\documentclass[11pt]{article}
\usepackage[top=20truemm,bottom=25truemm,left=23truemm,right=27truemm]{geometry}

\usepackage{amsmath,amssymb,amsfonts,graphics,graphicx,amscd,amsfonts,epsfig,color}
\usepackage{array, booktabs,bm,cite}
\usepackage{subfig}
\usepackage{here}

\newcommand {\beq}{\begin{equation}}
\newcommand {\eeq}{\end{equation}}
\newcommand {\beqa}{\begin{eqnarray}}
\newcommand {\eeqa}{\end{eqnarray}}

\newcommand {\Tr}{\mbox{Tr}}

\newcommand {\del}{\partial}

\newcommand {\expval}[1]{\left \langle #1 \right \rangle}

\newcommand {\comm}[2]{\boxbrackets{#1, #2}}
\newcommand {\pdv}[1]{\frac{\del}{\del #1}}

\newcommand {\boxbrackets}[1]{\left[ #1 \right]}
\numberwithin{equation}{section}

\allowdisplaybreaks

\begin{document}

\begin{titlepage}
\renewcommand{\thefootnote}{\fnsymbol{footnote}}
\begin{normalsize}
\begin{flushright}
\begin{tabular}{l}
April 2018
\end{tabular}
\end{flushright}
\end{normalsize}

~~\\

\vspace*{0cm}
    \begin{Large}
       \begin{center}
         {
         Equivalence of large-$N$ gauge theories on a group manifold \\ and its coset space
         }
       \end{center}
    \end{Large}
\vspace{1cm}

\begin{center}
           Kohta H{\sc atakeyama}\footnote
            {
e-mail address : 
hatakeyama.kohta.15@shizuoka.ac.jp}
           {\sc and}
           Asato T{\sc suchiya}\footnote
           {
e-mail address : tsuchiya.asato@shizuoka.ac.jp}\\
      \vspace{1cm}

{\it Department of Physics, Shizuoka University}\\
                {\it 836 Ohya, Suruga-ku, Shizuoka 422-8529, Japan}\\
         \vspace{0.3cm}     
 {\it Graduate School of Science and Technology, Shizuoka University}\\
               {\it 3-5-1 Johoku, Naka-ku, Hamamatsu 432-8011, Japan}

\end{center}

\hspace{5cm}


\vspace{0.5cm}

\begin{abstract}
\noindent
It was shown in arXiv:0912.1456 
that the large-$N$ reduction holds on group manifolds
in the sense that a large-$N$ gauge theory on a group manifold
is realized by a matrix model
which is obtained by dimensionally reducing the original theory to zero dimension.
In this note, generalizing the above statement,
we show that a large-$N$ gauge theory on a group manifold
is equivalent to a theory which is obtained by reducing the original theory to
its coset space. This is analogous to the statement of the large-$N$ reduction on
flat spaces that large-$N$ gauge theories are independent of the volume.
\end{abstract}
\vfill
\end{titlepage}
\vfil\eject

\setcounter{footnote}{0}

\section{Introduction}
\setcounter{equation}{0}
\renewcommand{\thefootnote}{\arabic{footnote}} 
The large-$N$ reduction \cite{EK} states that large-$N$ gauge theories are 
independent of the volume of the space-time on which they are defined.
(For further developments in the large-$N$ reduction, see \cite{Gross:1980br,Bhanot:1982sh,Parisi:1982gp,Gross:1982at,Das:1982ux,
GonzalezArroyo:1982hz, Eguchi:1982ta, GonzalezArroyo:1983ac, Narayanan:2003fc,Bietenholz:2006cz,Teper:2006sp,Kovtun:2007py,
Azeyanagi:2007su,Unsal:2008ch,Bringoltz:2008av,Ishii:2008ib,Bringoltz:2009kb,
Kawai:2009vb,Kawai:2010sf,GonzalezArroyo:2010ss,Lucini:2012gg}.)
In particular, this implies that large-$N$ gauge theories
are equivalent to the matrix models called the reduced models 
that are obtained by dimensionally reducing  
the original theories to zero dimension.
The large-$N$ reduction can give an efficient way to
define large-$N$ gauge theories nonperturbatively, and
is also conceptually interesting 
as examples of emergent space-time.
The large-$N$ reduction is also called the large-$N$ volume independence
when the aspect of volume independence is emphasized.

The large-$N$ reduction had been studied on flat space-time.
It was shown in \cite{Kawai:2009vb,Kawai:2010sf} 
that the large-$N$ reduction holds on group manifolds in the sense
that a large-$N$ gauge theory on a group manifold
is realized by a matrix model
which is obtained by reducing the original theory to zero dimension.

In this note, we examine whether phenomenon analogous to the large-$N$ volume independence  occurs on group manifolds.
 We find that it indeed does
in the sense that a large-$N$ gauge theory on a group manifold
$G$ is equivalent to the theory obtained by reducing it to a coset space $G/H$ 
where $H$ is a subgroup of $G$.
Here, we call this phenomenon the large-$N$ equivalence in dimensional reduction on group manifolds.

This paper is organized as follows.
In section 2, as a warm up, we examine the large-$N$ volume independence
on a torus.
In section 3, we review some properties of Lie groups and coset spaces.
In section 4, we study the large-$N$ equivalence in dimensional reduction on group manifolds.
Section 5 is devoted to conclusion and discussion.

\section{Large-$N$ volume independence on torus}
\setcounter{equation}{0}
In this section, as a warm-up, we examine the large-$N$ volume independence
on a $D$-dimensional torus $T^D \simeq U(1)^D$. We denote coordinates 
of $T^D$ by $x^{\mu}$ $(\mu=1,\dots,D)$, assuming, for simplicity, the periodicity
$x^{\mu} \sim x^{\mu}+L$. Using a positive integer $K$, we define a `reduced torus'
$T^D/(Z_K)^D$ whose coordinates are denoted by $\sigma^{\mu}$.
The periodicity for $\sigma^{\mu}$ is 
$\sigma^{\mu} \sim \sigma^{\mu}+l$ where $l=L/K$.
We have a relation
\begin{equation}
x^{\mu}=lu^{\mu}+\sigma^{\mu} 
\label{relation between x and y}
\end{equation}
with $u^\mu$ integers.

To illustrate the large-$N$ volume independence, 
we consider a scalar matrix field theory on $T^D$:
\begin{equation}
S = \int d^D x \; \Tr 
\left(\frac{1}{2} \del_\mu \phi(x) \del_\mu \phi(x)
+\frac{m^2}{2} \phi(x)^2 
+\frac{\kappa}{3} \phi(x)^3 \right) \ ,
\label{scalar field theory on torus}
\end{equation}
where $\phi(x)$ is a Hermitian matrix-valued field with the matrix size $N$.

We apply a following reduction rule to the above theory (\ref{scalar field theory on torus}):
\begin{equation}
\phi(x) \rightarrow e^{i P_\mu x^\mu} \phi(\sigma) e^{-i P_\mu x^\mu}
\quad
\mbox{with} \;\;\; P_\mu = 
\begin{pmatrix}
\frac{2\pi n_\mu^{(1)}}{L} & {} & {} \\
{} & \frac{2\pi n_\mu^{(2)}}{L} & {} \\
{} & {} & \ddots \\
\end{pmatrix}  \ ,\quad \int d^D x \rightarrow \frac{v}{v'}\int d^D\sigma  \ .
\label{reduction rule}
\end{equation}
Here the relation between $x^{\mu}$ and $\sigma^{\mu}$ is given by
(\ref{relation between x and y}), and
$P_{\mu}$ are constant diagonal matrices whose eigenvalues 
$2\pi n_{\mu}^{(i)}/L$ $(i=1,\dots,N)$ 
correspond to the momenta on $T^D$ distributed uniformly
in the momentum space. $v$ and $v'$ are given by
\begin{equation}
v=L^D/N \ , \quad v'=(2\pi/\Lambda)^D \ ,
\label{v and v'}
\end{equation}
where $\Lambda$ is a UV cutoff on $T^D/(Z_K)^D$.
(\ref{v and v'}) implies that $T^D$ is divided into $N$ cells 
with the volume of a unit cell given by $v$ and that 
$T^D/(Z_K)^D$ is divided into $l^D/v'$ cells with the volume of a unit cell given by
$v'$.
Then, we obtain the action of a reduced model defined on $T^D/(Z_K)^D$:
\begin{equation}
S_r = \frac{v}{v'} \int d^D \sigma \; \Tr 
\left(\frac{1}{2} \left(\del_{\sigma^\mu} \phi(\sigma)
+i\comm{P_\mu}{\phi(\sigma)} \right)^2 
+\frac{m^2}{2} \phi(\sigma)^2 
+\frac{\kappa}{3} \phi(\sigma)^3 \right) \ .
\label{reduced model}
\end{equation}
Note that $vl^D/v'$ can be viewed as an effective volume in the reduced model.

\begin{figure}
\begin{minipage}{0.47\hsize}
\centering
\includegraphics[width=3.5cm]{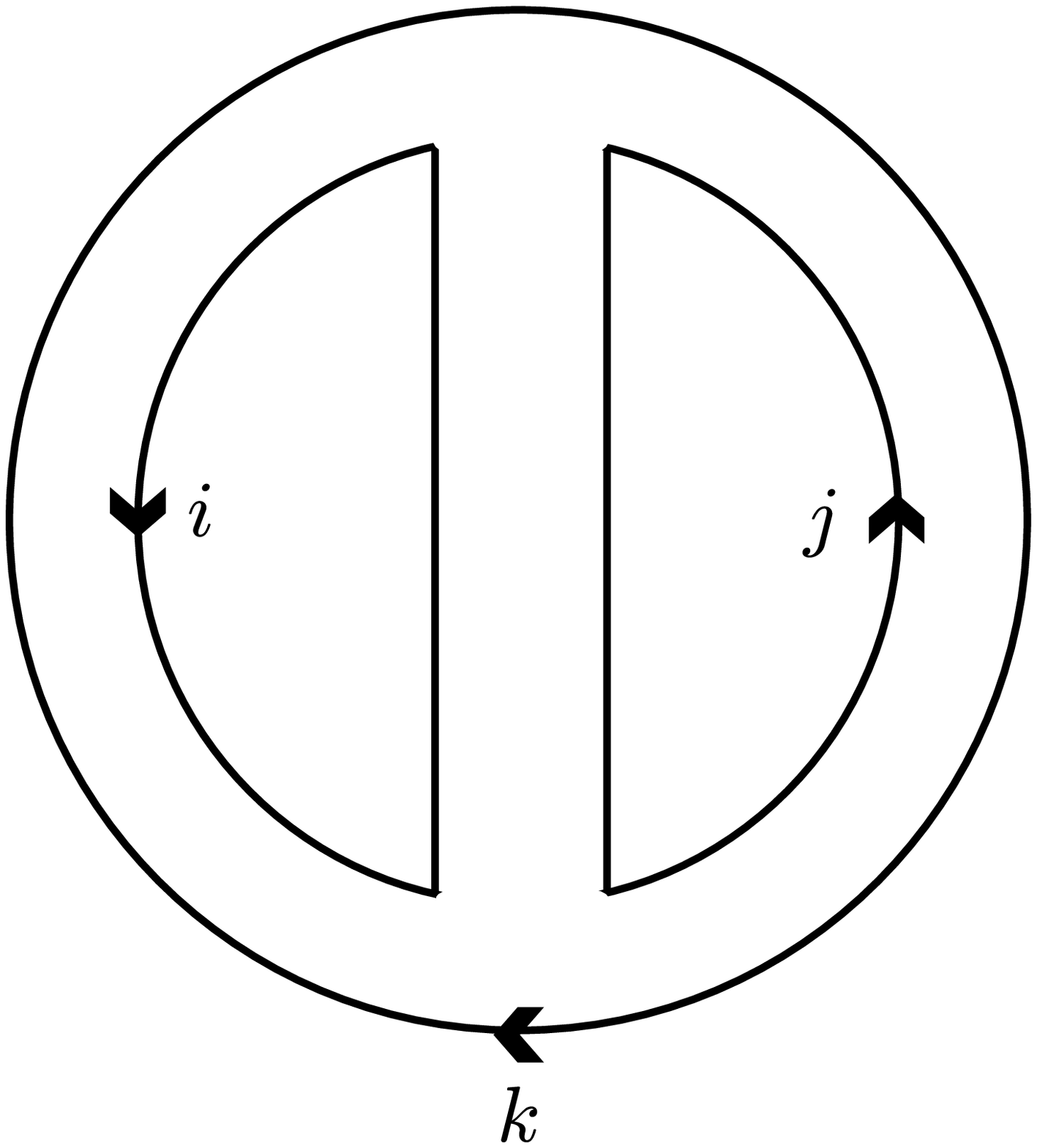}
\caption{Two-loop planar diagram for the free energy.}
\label{planar}
\end{minipage}
\begin{minipage}{0.1\hsize}
\centering
\phantom{aaa}
\end{minipage}
\begin{minipage}{0.47\hsize}
\vspace{0.15cm}
\centering
\includegraphics[width=4cm]{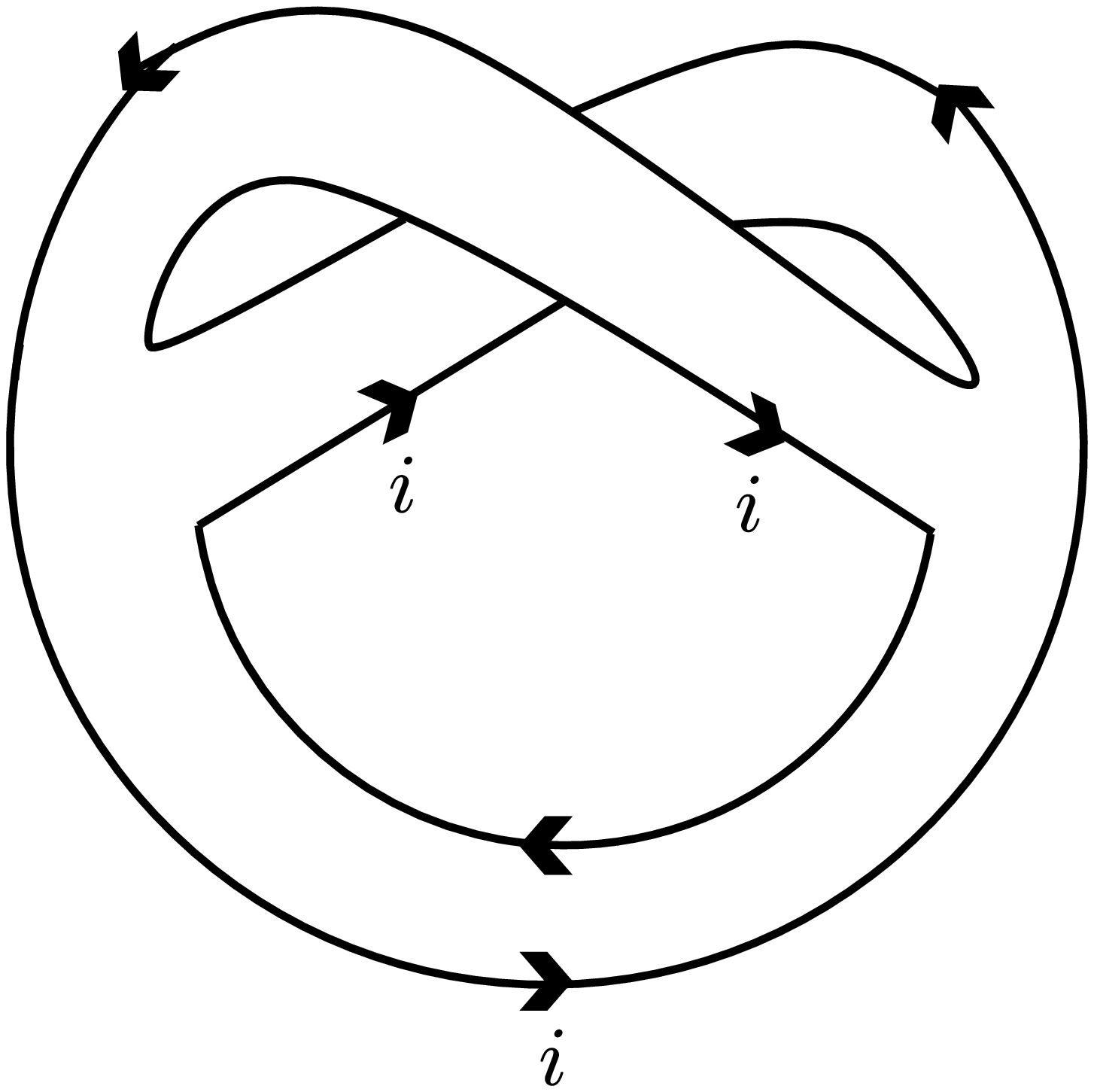}
\caption{Two-loop non-planar diagram for the free energy.}
\label{non-planar}
\end{minipage}
\end{figure}
We consider the two-loop contribution to the free energy.
There are two diagrams. One is planar (Fig.\ref{planar}) and the other non-planar (Fig.\ref{non-planar}).
First, we calculate them in the original theory (\ref{scalar field theory on torus}).
The planar diagram is calculated as
\begin{equation}
\frac{\lambda N^2}{6}\int d^D x d^D x' D(x-x')^3 \ ,
\label{planar diagram in original model}
\end{equation}
where $\lambda= \kappa^2 N$ is the 't Hooft coupling and $D(x-x')$ is the free propagator of 
the theory (\ref{scalar field theory on torus}) with $N=1$:
\begin{equation}
D(x-x') = \frac{1}{L^D}\sum_n
\frac{e^{i \frac{2\pi n_\mu}{L} \left(x_\mu-x'_\mu \right)}}{\left(\frac{2\pi n_{\mu}}{L} \right)^2 +m^2}  \ .
\end{equation}
The non-planar diagram is given by (\ref{planar diagram in original model})$/N^2$
so that it is suppressed by $1/N^2$ compared to the planar diagram in the 
$N\rightarrow\infty$ limit.

Next, we calculate them in the reduced model using a bilocal field representation for
matrices \cite{Kawai:2009vb,Kawai:2010sf}. We take a coordinate basis $|x\rangle$ in the vector space on which
$\phi(\sigma)$ and $P_{\mu}$ act and define a bilocal field 
\begin{equation}
\phi(\sigma,x,x')=\langle x | \phi(\sigma) |x'\rangle \ .
\end{equation}
The reduced model (\ref{reduced model}) is rewritten as 
\begin{align}
S_r &= \frac{v}{v'} \int d^D \sigma d^D x d^D x' \frac{1}{2} 
\phi(\sigma, x', x) \left[-\left(\del_{\sigma^\mu} + \del_\mu + \del'_\mu \right)^2 + m^2 \right] \phi(\sigma, x, x') \nonumber \\
&\quad 
+ \frac{\kappa v}{3 v'} \int d^D \sigma d^D x d^D x' d^D x'' \phi(\sigma, x, x') \phi(\sigma, x', x'') \phi(\sigma, x'', x) \ .
\end{align}

We make a change of variables, 
$
\bar{x}^{\mu}=x^{\mu} ,\ 
\tilde{x}^{\mu} = x^{\mu}-x'^{\mu} ,\ 
\bar{\sigma}^{\mu} = \sigma^{\mu} - x^{\mu}$,
which gives
\begin{equation}
\frac{\partial}{\partial \sigma^{\mu}}+\frac{\partial}{\partial x^{\mu}}
+\frac{\partial}{\partial x'^{\mu}}
=\frac{\partial}{\partial \bar{x}^{\mu}} \ .
\label{derivative}
\end{equation}
Thus, from (\ref{derivative}) we obtain the propagator
\begin{equation}
\expval{\phi(\sigma_1, x_1, x'_1) \phi(\sigma_2, x'_2, x_2)} = \frac{v'}{v} D(x_1 - x_2) 
\delta_L^{(D)}((x_1 - x'_1) - (x_2 - x'_2)) 
\delta_l^{(D)}((\sigma_1 - \sigma_2) - (x_1 - x_2)) \ ,
\end{equation}
where $\delta_L^{(D)}$ and $\delta_l^{(D)}$ are periodic delta functions with the
period $L$ and $l$, respectively.

\begin{figure}
\begin{minipage}{0.47\hsize}
\centering
\includegraphics[width=4.5cm]{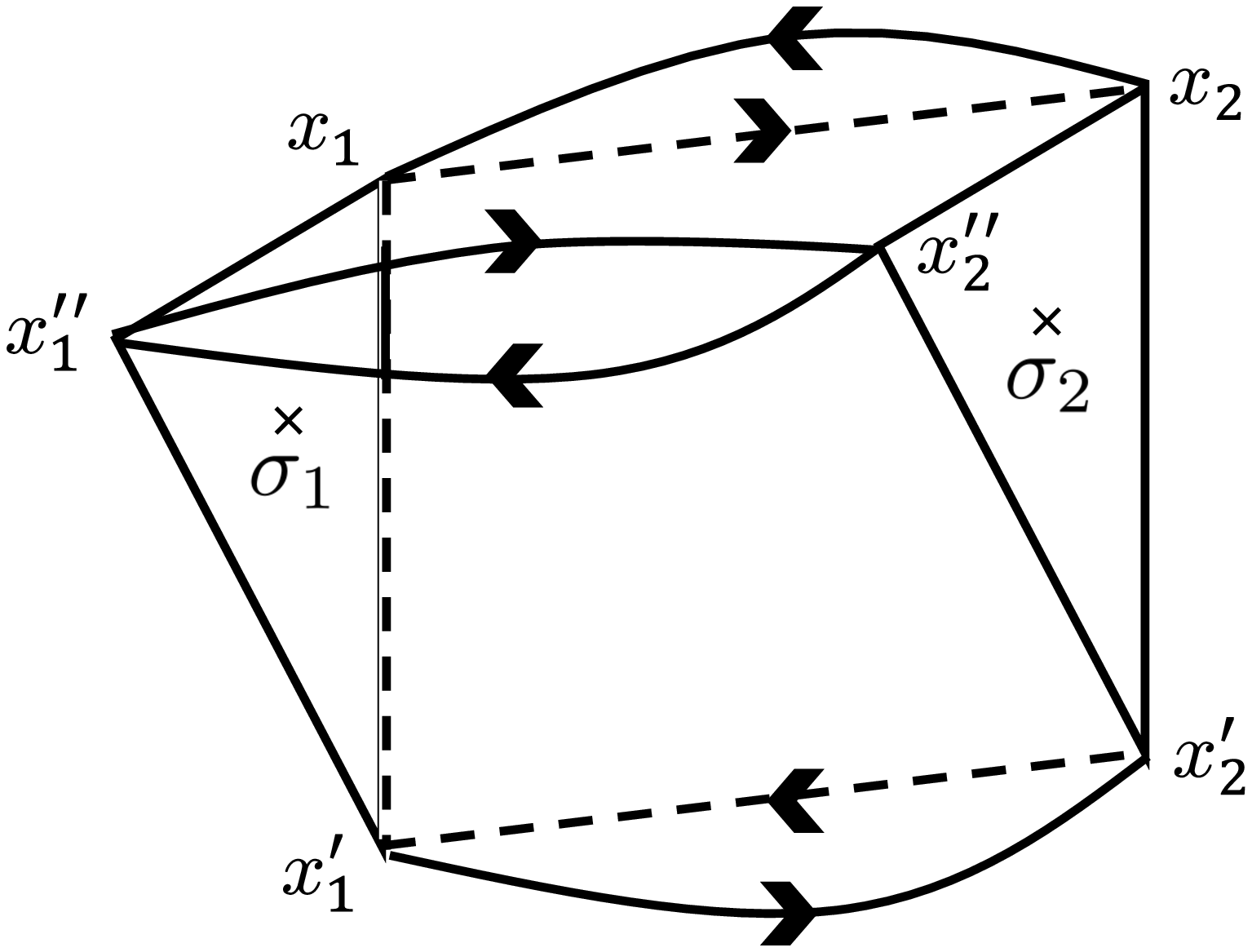}
\caption{Two-loop planar diagram in the bilocal representation for the free energy.}
\label{bi-local_planar}
\end{minipage}
\begin{minipage}{0.1\hsize}
\centering
\phantom{aaa}
\end{minipage}
\begin{minipage}{0.47\hsize}
\centering
\includegraphics[width=5cm]{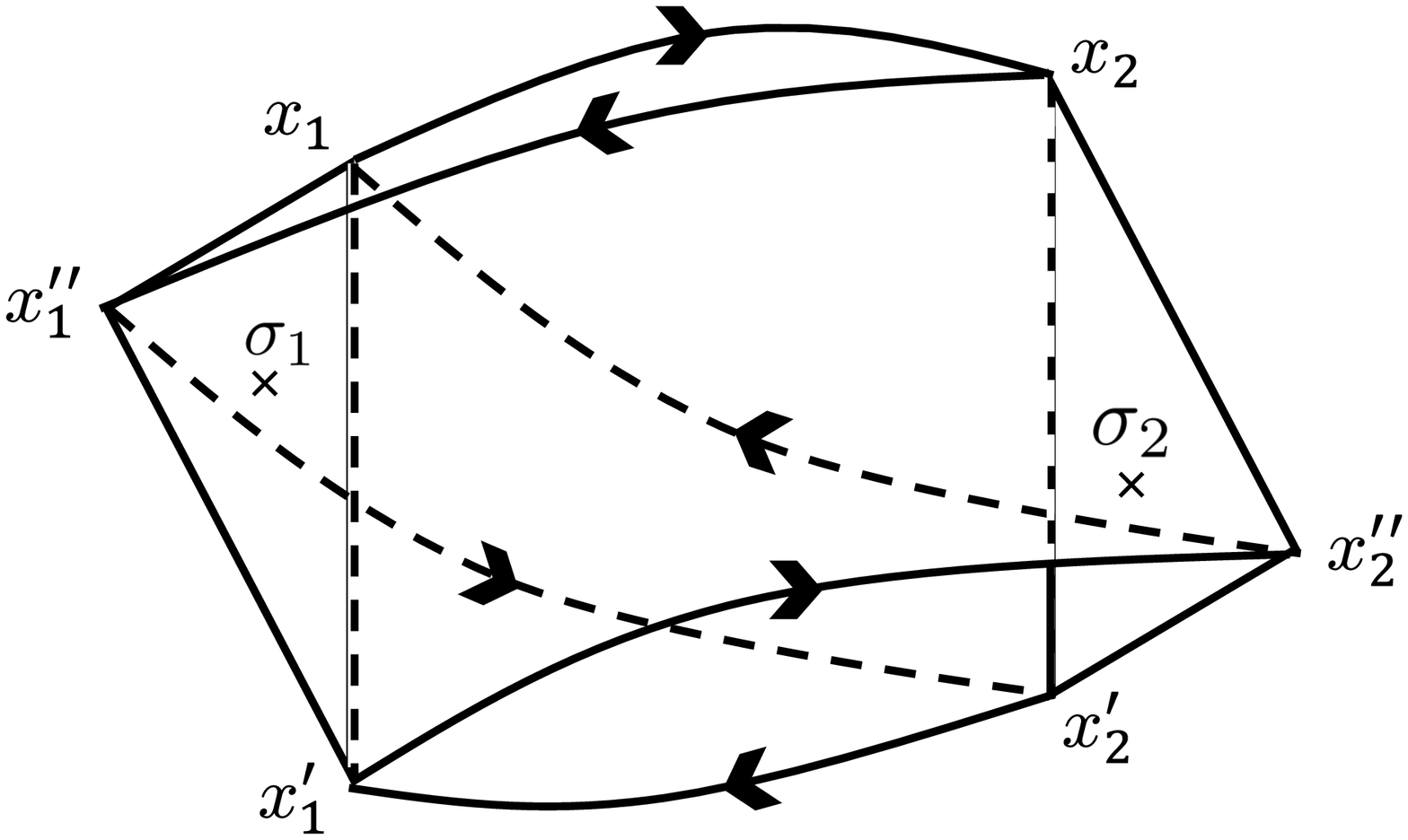}
\caption{Two-loop non-planar diagram in the bilocal representation for the free energy.}
\label{bi-local_non-planar}
\end{minipage}
\end{figure}
The planar diagram (Fig.\ref{bi-local_planar}) in the reduced model is calculated as
\begin{align}
&\quad 3 \cdot \frac{1}{2} \left(\frac{\kappa v}{3 v'} \right)^2 
\int d^D \sigma_1 d^D \sigma_2 d^D x_1 d^D x'_1 d^D x''_1 d^D x_2 d^D x'_2 d^D x''_2 
\nonumber \\
&\qquad \times \frac{v'}{v} D(x_1 - x_2) 
\delta_L^{(D)}((x_1 - x'_1) - (x_2 - x'_2)) 
\delta_l^{(D)}((\sigma_1 - \sigma_2) - (x_1 - x_2)) \nonumber \\
&\qquad \times \frac{v'}{v} D(x'_1 - x'_2) 
\delta_L^{(D)}((x'_1 - x''_1) - (x'_2 - x''_2)) 
\delta_l^{(D)}((\sigma_1 - \sigma_2) - (x'_1 - x'_2)) \nonumber \\
&\qquad \times \frac{v'}{v} D(x''_1 - x''_2) 
\delta_L^{(D)}((x''_1 - x_1) - (x''_2 - x_2)) 
\delta_l^{(D)}((\sigma_1 - \sigma_2) - (x''_1 - x''_2)) \nonumber \\
&= \frac{\kappa^2 v'}{6 v} \delta_L^{(D)}(0) L^{2D} 
\int d^D \sigma_1 d^D \sigma_2 d^D x_1 d^D x_2 D(x_1 - x_2)^3
\delta_l^{(D)}((\sigma_1 - \sigma_2) - (x_1 - x_2))^3 \nonumber \\
&= \frac{\kappa^2 v'}{6 v^2} L^{3D} 
\int d^D \sigma_1 d^D \sigma_2 d^D \tilde{x} D(\tilde{x})^3 
\delta_l^{(D)}((\sigma_1 - \sigma_2) - \tilde{x})^3 \nonumber \\
&= \frac{\kappa^2 L^{3D} v'}{6 v^2} \delta_l^{(D)}(0)^2 
\int d^D \sigma_1 d^D \sigma_2 \sum_u D(lu +\sigma_1 - \sigma_2)^3 \nonumber \\
&= \frac{\kappa^2 L^{3D}}{6 v^2 v'}  \int d^D \tilde{x} d^D \sigma_2  
D(\tilde{x} - \sigma_2)^3 \nonumber \\
&= \frac{\kappa^2 L^{3D} l^D}{6 v^2 v'}  \int d^D \tilde{x} D(\tilde{x})^3 
\nonumber \\
&= \frac{\kappa^2 l^D}{6 v^2 v'} (Nv)^{3} \int d^D \tilde{x} D(\tilde{x})^3  
\nonumber\\
&= \frac{vl^D}{v'}  \frac{\lambda N^2}{6}\frac{1}{L^D} \int d^D x d^D x' D(x - x')^3 \ ,
\label{planar diagram in bilocal reduced model}
\end{align}
where we have used $\delta_L(0)=1/v$ and $\delta_l(0)=1/v'$.

The non-planar diagram (Fig.\ref{bi-local_non-planar}) in the reduced model is calculated as
\begin{align}
&\quad 3 \cdot \frac{1}{2} \left(\frac{\kappa v}{3 v'} \right)^2 
\int d^D \sigma_1 d^D \sigma_2 d^D x_1 d^D x'_1 d^D x''_1 d^D x_2 d^D x'_2 d^D x''_2 
\nonumber \\
&\qquad \times \frac{v'}{v} D(x_1 - x_2) 
\delta_L^{(D)}((x_1 - x'_1) - (x_2 - x'_2)) 
\delta_l^{(D)}((\sigma_1 - \sigma_2) - (x_1 - x_2)) \nonumber \\
&\qquad \times \frac{v'}{v} D(x'_1 - x''_2) 
\delta_L^{(D)}((x'_1 - x''_1) - (x''_2 - x_2)) 
\delta_l^{(D)}((\sigma_1 - \sigma_2) - (x'_1 - x''_2)) \nonumber \\
&\qquad \times \frac{v'}{v} D(x''_1 - x'_2) 
\delta_L^{(D)}((x''_1 - x_1) - (x'_2 - x''_2)) 
\delta_l^{(D)}((\sigma_1 - \sigma_2) - (x''_1 - x'_2)) \nonumber \\
&= \frac{\kappa^2 v'}{6 v} \delta_L^{(D)}(0) 
\int d^D \sigma_1 d^D \sigma_2 d^D x_1 d^D x'_1 d^D x_2 d^D x''_2 
D(x_1 - x_2) D(x'_1 - x''_2) D(x_1 - x''_2) \nonumber \\
& 
\qquad \times 
\delta_l^{(D)}((\sigma_1 - \sigma_2) - (x_1 - x_2)) 
\delta_l^{(D)}((\sigma_1 - \sigma_2) - (x'_1 - x''_2)) 
\delta_l^{(D)}((\sigma_1 - \sigma_2) - (x_1 - x''_2))  \nonumber \\
&= 
\frac{\kappa^2 v'}{6 v^2} L^{D}
\int d^D \sigma_1 d^D \sigma_2 \sum_{u, u', u''} D(lu +\sigma_1 - \sigma_2) 
D(lu' +\sigma_1 - \sigma_2) D(lu'' +\sigma_1 - \sigma_2) \nonumber\\
&= 
\left\{\frac{v l^D}{v'}\frac{\lambda N^2}{6}\frac{l^D}{L^{2D}}
\int d^D \sigma_1 d^D \sigma_2 \sum_{u, u', u''} 
D(lu +\sigma_1 - \sigma_2) D(lu' +\sigma_1 - \sigma_2) D(lu'' +\sigma_1 - \sigma_2)
\right\} \times \left(\frac{v'}{l^D}\right)^2 \ . \nonumber\\
\label{non-planar diagram in bilocal reduced model}
\end{align}
We see again that the non-planar diagram is suppressed compared to the 
planar diagram in the $v'\rightarrow 0$ limit, because the quantity in the curly
bracket in (\ref{non-planar diagram in bilocal reduced model}) 
has the same order of magnitude as 
(\ref{planar diagram in bilocal reduced model}).

The non-planar diagram is suppressed compared to the planar diagram
in both the original and reduced models. By comparing the planar contribution
(\ref{planar diagram in original model}) and (\ref{planar diagram in bilocal reduced model}),
we find a relation between the free energy $F$ in the original model and
the one $F_r$ in the reduced model in the $N\rightarrow\infty$ limit:
\begin{equation}
\frac{F}{N^2 V} = \frac{F_r}{N^2 v V'/v'} \ ,
\label{relation for free energy}
\end{equation}
where $V=L^D$ and $V'=l^D$ 
are the volumes of $T^D$ and $T^D/(Z_K)^D$, respectively, 
and the LHS and RHS correspond to the planar contribution to
the free energy per unit volume divided by $N^2$ in the original and reduced
models, respectively.
In a similar manner, by referring the argument in \cite{Gross:1982at}, one can show 
that the relation (\ref{relation for free energy}) holds 
to all orders in perturbative expansion. 
It is also easy to show 
a correspondence between correlation functions in the $N\rightarrow\infty$ limit\cite{Gross:1982at}:
\begin{equation}
\frac{1}{N^{q/2+1}} \expval{\Tr (\phi(x_1) \phi(x_2) \cdots \phi(x_q))} 
=\frac{1}{N^{q/2+1}} \expval{\Tr \left(\hat{\phi}(x_1) \hat{\phi}(x_2) 
\cdots \hat{\phi}(x_q) \right)}_r \ ,
\label{relation for correlation functions}
\end{equation}
where $\langle \cdots \rangle$ and $\langle \cdots \rangle_r$ stand for
the expectation value in the original and reduced models, respectively, and
$\hat{\phi}(x) = e^{iP_\mu x^\mu} \phi(\sigma) e^{-iP_\mu x^\mu}$ with 
(\ref{relation between x and y}).
Thus, we find that the large-$N$ volume independence holds on a torus
in the sense that a theory on $T^D$ is equivalent to a certain theory 
on $T^D/(Z_K)^D$ in the large-$N$ limit.

Finally, we consider Yang-Mills theory on $T^D$:
\begin{equation}
S=\frac{1}{4\kappa^2}\int d^Dx \; \mbox{Tr}(F_{\mu\nu}F_{\mu\nu}) \ ,
\label{YM on torus}
\end{equation}
where $F_{\mu\nu}=\partial_\mu A_\nu - \partial_\nu A_\mu +i[A_\mu,A_\nu]$.
By applying the reduction rule (\ref{reduction rule}) to (\ref{YM on torus}),
we obtain 
\begin{equation}
S_r = \frac{v}{v'}\frac{1}{4\kappa^2}\int d^{D}\sigma \; 
\mbox{Tr} \left(\tilde{F}_{\mu\nu}\tilde{F}_{\mu\nu} \right) \ ,
\label{reduce model on small torus}
\end{equation}
where $\tilde{F}_{\mu\nu}=\partial_{\sigma^\mu} \tilde{A}_\nu 
- \partial_{\sigma^\nu} \tilde{A}_\mu 
+i \left[\tilde{A}_\mu,\tilde{A}_\nu \right]$ 
with $\tilde{A}_{\mu}(\sigma)=P_{\mu}+A_{\mu}(\sigma)$.
Namely, the reduced model agrees with the one that is obtained by dimensionally
reducing the original model to $T^D/(Z_K)^D$.
If the background $\tilde{A}_{\mu}=P_{\mu}$ is stable in the reduced model
(\ref{reduce model on small torus}), the reduced model is equivalent to
the original model (\ref{YM on torus}) in the $N\rightarrow\infty$ with
$\kappa^2N$ fixed in the sense that
(\ref{relation for free energy}) holds and a following relation for Wilson loops
also holds:
\begin{equation}
\left\langle \frac{1}{N}P \exp \left(i\int_0^1d\zeta 
\frac{dx^{\mu}(\zeta)}{d\zeta} A_{\mu}(x(\zeta)) \right) \right\rangle
=\left\langle  \frac{1}{N}P \exp \left(i\int_0^1d\zeta 
\frac{dx^{\mu}(\zeta)}{d\zeta} \tilde{A}_{\mu}(\sigma(\zeta)) \right) \right\rangle_r \ ,
\label{relation for Wilson loop}
\end{equation}
where $x^{\mu}(\zeta)$ and $\sigma^{\mu}(\zeta)$ are related as
(\ref{relation between x and y}).
Namely, the large-$N$ volume independence holds literally.
Note that the stability depends on the dynamics of the model\footnote{
Instability corresponds to SSB of
the so-called $U(1)^D$ symmetry  or the center invariance.}.

\section{Group manifolds and coset spaces}
\setcounter{equation}{0}

In this section, we review some basic facts about group manifolds and coset spaces.
For more details, see, for instance, \cite{Ishii:2008tm,Kawai:2009vb,Kawai:2010sf}.
Let $G$ be a compact simply connected Lie group\footnote{If $G$ is not simply connected, the reduced model is not obtained in a globally consistent way.} and $H$ be a Lie subgroup of $G$.
$D$ and $d$ denote the dimensions of $G$ and $H$, respectively.
Then, the dimension of $G/H$ is $D-d$.
$x^M$ $(M=1,\dots,D)$, $y^{m}$ $(m=D-d+1,\dots,D)$ and
$\sigma^ {\mu}$ $(\mu=1,\dots,D-d)$ 
denote the coordinates of $G$, $H$ and $G/H$, respectively, while
$A,B=1,\dots,D$, $a,b=D-d+1,\dots,D$ and $\alpha,\beta=1,\dots,D-d$ are
the corresponding local Lorentz indices.

Let $t_A$ be a basis for the Lie algebra of $G$ in which $t_a$ are a basis for the Lie algebra of $H$.
$t_A$ satisfy commutation relations $[t_A,t_B]=if_{ABC}t_C$ with $f_{ABC}$ completely anti-symmetric and $f_{ab\alpha}=0$.
$g(x)\in \:G$ is factorized locally as $g(x)=L(\sigma)h(y)$, where $h(y)\in\:H$.
The isometry of $G$ is the $G\times G$ symmetry, 
where one corresponds to the left translation and the other 
the right translation. 
Only the left translation survives as the isometry of $G/H$. 

A $D\times D$ matrix $Ad(g)$ for $g \in G$ is defined by $g\: t_A\: g^{-1}=t_B\: Ad(g)_{BA}$.
It is easy to show that $Ad(g)_{AB}Ad(g)_{AC}=\delta_{BC}$.
Note that if $h$ is an element of $H$,
$
Ad(h)_{\alpha a}=Ad(h)_{a\alpha}=0
$, 
which implies that 
$
Ad(h)_{\alpha\beta}Ad(h)_{\alpha\gamma}=\delta_{\beta\gamma}, \ 
Ad(h)_{ab}Ad(h)_{ac}=\delta_{bc}.
$

The right invariant 1-form $E^A_M$ and the left invariant 1-form $S^A_M$ 
are defined by
\begin{equation}
\partial_M g(x)g^{-1}(x)=-iE^A_M(x)\:t_A \ , \quad g^{-1}(x)\partial_M g(x) = i S^A_M(x) \: t_A \ .
\end{equation}
They satisfy the Maurer-Cartan equation
\begin{equation}
\partial_ME^A_N-\partial_NE^A_M-f_{ABC}E^B_ME^C_N=0 \ , \quad
\partial_MS^A_N-\partial_NS^A_M-f_{ABC}S^B_MS^C_N=0 \ .
\label{Maurer-Cartan equation}
\end{equation}
Defining $e^A_{\mu}$, $\tilde{e}^a_m$, $s^A_{\mu}$ and $\tilde{s}^a_m$  by
\begin{align}
\partial_{\mu}L(\sigma)L^{-1}(\sigma)=-ie^A_{\mu}(\sigma)\:t_A \ &, \quad
\partial_mh(y)h^{-1}(y)=-i\tilde{e}^a_m(y)\:t_a \ , \nonumber\\
L^{-1}(\sigma)\del_\mu L(\sigma) = i s_\mu^A(\sigma) t_A \ &, \quad
h^{-1}(y) \del_m h(y) = i \tilde{s}_m^a(y) t_a \ ,
\end{align}
we obtain the relations:
\begin{align}
&E^{\alpha}_{\mu}(x)=e^{\alpha}_{\mu}(\sigma) \ , \quad
E^a_{\mu}(x)=e^{a}_{\mu}(\sigma) \ , \quad
E^{\alpha}_m(x)=Ad(L)_{\alpha b}(\sigma)\tilde{e}^b_m(y) \ , \quad
E^a_m(x)=Ad(L)_{ab}(\sigma)\tilde{e}^b_m(y) \ , \nonumber\\
&S_\mu^\alpha(x) = Ad(h^{-1})_{\alpha \beta}(y) s_\mu^\beta(\sigma) \ , 
\quad
S_\mu^a(x) = Ad(h^{-1})_{ab}(y) s_\mu^b(\sigma) \ , \quad
S_m^\alpha(x) = 0 \ , \quad
S_m^a(x) = \tilde{s}_m^a(y) \ .
\label{components of E}
\end{align}

A metric of $G$, 
$
G_{MN}=E^A_ME^A_N =S^A_MS^A_N
$, 
is right and left invariant.
By using (\ref{components of E}), we obtain
$
ds_G^2= s_\mu^\beta s_\nu^\beta d\sigma^\mu d\sigma^\nu 
+ \left(Ad(h^{-1})_{ba} \tilde{s}_m^b dy^m + s_\mu^a d\sigma^\mu \right)^2
$,
where  the invariant metric of $G/H$, $g_{\mu\nu}$, is given by
$
g_{\mu\nu}=s^{\alpha}_{\mu}s^{\alpha}_{\nu}
$.
The Haar measure of $G$ is given by
$
dg=d^Dx \sqrt{G(x)}
$, 
where $G(x)$ is $\det G_{MN}(x)$.
It is factorized as
$
dg=d^{D-d}\sigma d^dy \sqrt{g(\sigma)} \det\tilde{s}^a_m(y)
$.
We denote the invariant measure of $G/H$, $d^{D-d}\sigma\sqrt{g}$, by $dL$.

We define the right invariant Killing vectors ${\cal L}_A$ and
the left invariant Killing vectors ${\cal R}_A$ by
\begin{equation}
{\cal L}_A=-iE^M_A\frac{\partial}{\partial x^M} \ , \quad
\mathcal{R}_A = -i S_A^M \pdv{x^M} \ ,
\label{right invariant Killing vector}
\end{equation}
where $E^M_A$ and $S^M_A$ are the inverses of $E^A_M$ and $S^A_M$, respectively.
${\cal L}_A$ and ${\cal R}_A$ generate the left translation and right translation,
respectively, and obey the following commutation relations:
\begin{equation}
[{\cal L}_A,{\cal L}_B]=i f_{ABC}{\cal L}_C, \;\;\;
[{\cal R}_A,{\cal R}_B]=i f_{ABC}{\cal R}_C, \;\;\;
[{\cal L}_A,{\cal R}_B]=0 \ .
\end{equation}
By using (\ref{components of E}), we obtain
\begin{align}
&{\cal L}_{\alpha}=i s^{\mu}_{\beta}Ad(L)_{\beta\alpha}
\frac{\partial}{\partial \sigma^{\mu}}
-i\tilde{e}^m_b \left(Ad(L)_{b\alpha}-s^b_{\rho}s^{\rho}_{\beta}Ad(L)_{\beta\alpha} \right)
\frac{\partial}{\partial y^m} \ , \nonumber\\
&{\cal L}_a = i s^{\mu}_{\beta}Ad(L)_{\beta a}\frac{\partial}{\partial \sigma^{\mu}}
-i\tilde{e}^m_b \left(Ad(L)_{ba}-s^b_{\rho}s^{\rho}_{\beta}Ad(L)_{\beta a} \right)
\frac{\partial}{\partial y^m} \ , \nonumber\\
&\mathcal{R}_\alpha = -i Ad(h^{-1})_{\alpha \beta} s_\beta^\mu \pdv{\sigma^\mu}
+ i  Ad(h^{-1})_{bc} Ad(h^{-1})_{\alpha \beta}\:\tilde{s}_b^m 
s_\nu^c s_\beta^\nu \pdv{y^m} \ , \nonumber\\
&\mathcal{R}_a = -i \tilde{s}_a^m \pdv{y^m} \ ,
\label{L_a and L_alpha}
\end{align}
where $s^{\mu}_{\alpha}$, $\tilde{s}^m_a$ and $\tilde{e}^m_a$ are the inverses of 
$s^{\alpha}_{\mu}$, $\tilde{s}^a_m$ and $\tilde{e}^a_m$, respectively.
${\cal L}'_A=is^{\mu}_{\beta}Ad(L)_{\beta A}\frac{\partial}{\partial \sigma^{\mu}}$
are the Killing vectors on $G/H$ and are indeed independent of $y$.

Let us consider a scalar matrix field theory on $G$ given by\footnote{The invariant 1-forms, the Killing vectors, the integral measures and the delta functions are all well-defined globally. Hence, all expressions on group manifolds and their coset spaces make sense globally, although they look dependent on coordinate patches.}
\begin{align}
S &= \int d^D x \sqrt{G(x)} \; \mbox{Tr}
\left[\frac{1}{2} G^{MN} \del_M \phi(x) \del_N \phi(x) +\frac{m^2}{2} \phi(x)^2 +\frac{\kappa}{3} \phi(x)^3 \right] \nonumber \\
&= \int d^{D-d} \sigma d^d y \sqrt{g(\sigma)} \det \tilde{s}_m^a(y)
\mbox{Tr} \left[-\frac{1}{2} \left(\mathcal{L}_A \phi(x) \right)^2 
+\frac{m^2}{2} \phi(x)^2 +\frac{\kappa}{3} \phi(x)^3 \right] \ ,
\label{scalar field theory on G}
\end{align}
where $\phi(x)$ is an $N\times N$ hermitian matrix.
This theory has the $G\times G$ symmetry.
Namely, (\ref{scalar field theory on G}) is invariant under
$\delta \phi = \epsilon {\cal L}_A \phi$ or $\delta \phi = \epsilon {\cal R}_A \phi$.
We impose a constraint
$
{\cal R}_a \phi(x) = 0
$, 
which implies from (\ref{L_a and L_alpha}) that $\phi$ 
is independent of $y$. Then, the theory 
(\ref{scalar field theory on G}) is reduced to the theory on $G/H$ as
\begin{align}
S&=\int d^{D-d} \sigma d^d y \sqrt{g} \det \tilde{s}_m^a
\mbox{Tr}\left[-\frac{1}{2} \left( {\cal L}'_A\phi \right)^2 
+\frac{m^2}{2} \phi^2 +\frac{\kappa}{3} \phi^3 \right] \nonumber \\
&= V_{H} \int d^{D-d} \sigma \sqrt{g} \;
\mbox{Tr}\left(\frac{1}{2} g^{\mu \nu} \del_\mu \phi \del_\nu \phi 
+\frac{m^2}{2} \phi^2 +\frac{\kappa}{3} \phi^3 \right)  \ ,
\label{scalar field theory on G/H}
\end{align}
where $V_H$ is the volume of $H$. The theory (\ref{scalar field theory on G/H})
has the left $G$ symmetry.
Note that this is a consistent truncation in the sense that
every solution to the equation of motion in (\ref{scalar field theory on G/H}) is 
also a solution to the equation of motion in (\ref{scalar field theory on G}).

As an example, 
we consider $SU(2) \simeq S^3$ and $SU(2)/U(1) \simeq S^2$.
We have
\begin{equation}
g = e^{-i\varphi \sigma_3/2} e^{-i\theta \sigma_2/2} e^{-i\psi \sigma_3/2} \ , \quad
L = e^{-i\varphi \sigma_3/2} e^{-i\theta \sigma_2/2} \ , \quad
h = e^{-i\psi \sigma_3/2} \ ,
\end{equation}
where $\theta$, $\varphi$ and $\psi$ are the Euler angles,
and $\sigma_i \; (i=1,2,3)$ are the Pauli matrices.
Here $\mu = (\theta, \varphi),\ m = \psi,\ \alpha =(1, 2)$, and $a = 3$.
${\cal L}_A$ are given by
\begin{align}
&{\cal L}_1 =  -i\left(-\sin\varphi \frac{\partial}{\partial {\theta}}
-\cot\theta\cos\varphi\frac{\partial}{\partial \varphi}
+\frac{\cos\varphi}{\sin\theta}\frac{\partial}{\partial \psi}\right)  \ , \nonumber\\
&{\cal L}_2 =  -i\left(-\cos\varphi \frac{\partial}{\partial {\theta}}
-\cot\theta\sin\varphi\frac{\partial}{\partial \varphi}
+\frac{\sin\varphi}{\sin\theta}\frac{\partial}{\partial \psi}\right)  \ , \nonumber\\
&{\cal L}_3 = -i\frac{\partial}{\partial \varphi}  \ .
\end{align}
${\cal R}_A$ are given by
\begin{align}
\mathcal{R}_1 
&= -i \left(-\sin \psi \frac{\partial}{\partial \theta} + \frac{\cos \psi}{\sin \theta} \frac{\partial}{\partial \varphi} -\cot \theta \cos \psi \frac{\partial}{\partial \psi} \right) \ , \nonumber \\
\mathcal{R}_2 
&= -i\left(-\cos \psi \frac{\partial}{\partial \theta} - \frac{\sin \psi}{\sin \theta} \frac{\partial}{\partial \varphi} 
+\cot \theta \sin \psi \frac{\partial}{\partial \psi} \right) \ , \nonumber \\
\mathcal{R}_3 
&= i \frac{\partial}{\partial \psi} \ .
\end{align}
The right and left invariant metric of $SU(2)$ is given by
$ds^2 = d\theta^2 + \sin^2\theta d\varphi^2 +(d\psi + \cos\theta d\varphi)^2
$.
The first and second terms in the RHS give the metric of $SU(2)/U(1)$.
The Haar measure takes the form
$
dg = \sin\theta d\theta d\varphi d\psi 
$.

\section{Bilocal representation for the reduced model on $G/H$}
\setcounter{equation}{0}
We consider a coordinate basis $|g\rangle$ for $G$ as in the case of torus.
We define the generators of the left translation $\hat{L}_A$ by
$
e^{i\epsilon \hat{L}_A} |g\rangle = |e^{i\epsilon t_A}g\rangle
$\footnote{$\hat{L}_A$ are the generators in the regular representation \cite{Kawai:2009vb,Kawai:2010sf}.}.
It is easy to show that
$
\hat{L}_A|g\rangle = -{\cal L}_A |g\rangle,  \  
\langle g|\hat{L}_A = {\cal L}_A \langle g| 
$.
We denote the volumes of $G$ and $G/H$ by $V$ and $V'$, respectively.
To obtain a reduced model
of (\ref{scalar field theory on G}) defined on $G/H$,
we apply the following rule
\begin{equation}
\phi(g) \rightarrow \phi(L) \ ,  \quad
{\cal L}_A \rightarrow \left[\hat{L}_A,\; \right] \ , \quad
\int dg \rightarrow \frac{v}{v'}\int dL=\frac{v}{v'}\int d^{D-d}\sigma \sqrt{g} \ .
\end{equation}
The first rule is realized by imposing 
$
{\cal R}_a \phi =0
$.
Thus, by introducing the bilocal representation for $\phi(L)$
\begin{equation}
\phi(L,g,g') = \langle g | \phi(L) | g'\rangle \ ,
\end{equation}
we obtain a bilocal representation of the reduced model:
\begin{align}
S_r &= \frac{v}{v'} \int dL dg dg' \frac{1}{2} \phi(L, g', g) 
\left [\left({{\cal L}'}^L_A+{\cal L}^g_A+{\cal L}^{g'}_A\right)^2 
+m^2 \right ] \phi(L, g, g')
\nonumber \\
&\qquad + \frac{\kappa v}{3v'} \int dL dg dg' dg'' \phi(L, g, g') \phi(L, g', g'') \phi(L, g'', g) \ .
\end{align}
As in the case of a torus, we make a change of variables 
$
w=g , \  \xi = g'^{-1}g , \  \rho=g^{-1}L 
$
and obtain a relation
\begin{equation}
\left({{\cal L}'}^L_A+{\cal L}^g_A+{\cal L}^{g'}_A \right)\phi(L,g,g') ={\cal L}^w_A\phi(L,g,g') \ .
\end{equation}
Thus, the propagator is read off as
\begin{equation}
\expval{\phi(L_1, g_1, g'_1)\phi(L_2, g'_2, g_2)} 
= \frac{v'}{v}\Delta \left(g_1 g_2^{-1} \right) 
\delta \left({g'_1}^{-1} g_1, {g'_2}^{-1} g_2 \right)
\delta_{G/H} \left(g_1^{-1} L_1, g_2^{-1} L_2 \right) \ ,
\end{equation}
where $\Delta(g_1g_2^{-1})$ is the propagator of the original model 
(\ref{scalar field theory on G}) with $N=1$, $\delta(g_1,g_2)$ is the right and left
invariant delta function on $G$, and $\delta_{G/H}(L_1,L_2)$ is the left invariant
delta function on $G/H$.

We consider the two-loop contribution to the free energy in the original and reduced models.
The planar diagram (Fig.\ref{bi-local_planar}) in the original theory is calculated as
\begin{equation}
\frac{\lambda N^2}{6} \int dg_1 dg_2 \Delta(g_1g_2^{-1})^3 \ .
\label{planar on G}
\end{equation}
The non-planar diagram (Fig.\ref{bi-local_non-planar}) is calculated as (\ref{planar on G})$/N^2$.
The planar diagram (Fig.\ref{bi-local_planar}) in the reduced model is calculated as
\begin{align}
&
3 \cdot \frac{1}{2} \left(\frac{\kappa v}{3v'} \right)^2 \left(\frac{v'}{v} \right)^3 
\int dL_1 dL_2 dg_1 dg'_1 dg''_1 dg_2 dg'_2 dg''_2 \nonumber \\
&\qquad \times
\Delta \left(g_1 g_2^{-1} \right) \delta \left({g'_1}^{-1} g_1, {g'_2}^{-1} g_2 \right)
\delta_{G/H} \left(g_1^{-1} L_1, g_2^{-1} L_2 \right) \nonumber \\
&\qquad \times
\Delta \left(g'_1 {g'_2}^{-1} \right) \delta \left({g''_1}^{-1} g'_1, {g''_2}^{-1} g'_2 \right)
\delta_{G/H} \left({g'_1}^{-1} L_1, {g'_2}^{-1} L_2 \right) \nonumber \\
&\qquad \times
\Delta \left(g''_1 {g''_2}^{-1} \right) \delta \left({g_1}^{-1} g''_1, {g_2}^{-1} g''_2 \right)
\delta_{G/H} \left({g''_1}^{-1} L_1, {g''_2}^{-1} L_2 \right) \ .
\end{align}
A change of variables $g'_2 = g_2 g_1^{-1}g'_1$ and $g''_2 = g_2 g_1^{-1} g''_1$
leads to 
\begin{align}
& \frac{\kappa^2 v'}{6v} \delta(0) \int dL_1 dL_2 dg_1 dg'_1 dg''_1 dg_2 
\Delta \left(g_1 g_2^{-1} \right)^3 \delta_{G/H} \left(g_1^{-1} L_1, g_2^{-1} L_2 \right) \nonumber \\
&\qquad \times
\delta_{G/H} \left({g'_1}^{-1} L_1, {g'_1}^{-1} g_1 g_2^{-1} L_2 \right) 
\delta_{G/H} \left({g''_1}^{-1} L_1, {g''_1}^{-1} g_1 g_2^{-1} L_2 \right)    \ .
\end{align}
Making a further change of variables 
${g'_1}^{-1} g_1 \rightarrow {g'_1}^{-1}$ and ${g''_1}^{-1} g_1 \rightarrow {g''_1}^{-1}$, 
we obtain
\begin{align}
&\quad \frac{\kappa^2 v'}{6v} \delta(0) \int dL_1 dL_2 dg_1 dg'_1 dg''_1 dg_2 
\Delta \left(g_1 g_2^{-1} \right)^3 \delta_{G/H} \left(g_1^{-1} L_1, g_2^{-1} L_2 \right) \nonumber\\
&\quad \qquad \times 
\delta_{G/H} \left({g'_1}^{-1} g_1^{-1} L_1, {g'_1}^{-1} g_2^{-1} L_2 \right) 
\delta_{G/H} \left({g''_1}^{-1} g_1^{-1} L_1, {g''_1}^{-1} g_2^{-1} L_2 \right) \nonumber \\
&= \frac{\kappa^2 v'}{6v} \delta(0) \int dL_1 dL_2 dg_1 dg'_1 dg''_1 dg_2 
\Delta \left(g_1 g_2^{-1} \right)^3 \delta_{G/H} 
\left(g_1^{-1} L_1, g_2^{-1} L_2 \right)^3 \nonumber \\
&= \frac{\kappa^2 v'}{6v} \delta(0) {V}^2 \delta_{G/H}(0)^2 V'
\int dg_1 dg_2 \Delta \left(g_1 g_2^{-1} \right)^3 \nonumber \\
&= \frac{vV'}{v'} \frac{\lambda N^2}{6 V} 
\int dg_1 dg_2 \Delta \left(g_1 g_2^{-1} \right)^3 \ .
\label{planar on G/H}
\end{align}
In the above calculation $\delta(0)=1/v,\ \delta_{G/H}(0)=1/v',
\ V=Nv$ and $\lambda=\kappa^2 N$ have been used.
The non-planar diagram (Fig.\ref{bi-local_non-planar}) in the reduced model is calculated as
\begin{align}
&
3 \cdot \frac{1}{2} \left(\frac{\kappa v}{3v'} \right)^2 \left(\frac{v'}{v} \right)^3 
\int dL_1 dL_2 dg_1 dg'_1 dg''_1 dg_2 dg'_2 dg''_2 \nonumber \\
&\quad \times 
\Delta \left(g_1 g_2^{-1} \right) \delta \left({g'_1}^{-1} g_1, {g'_2}^{-1} g_2 \right)
\delta_{G/H} \left(g_1^{-1} L_1, g_2^{-1} L_2 \right) \nonumber \\
&\quad \times
\Delta \left(g'_1 {g''_2}^{-1} \right) \delta \left({g''_1}^{-1} g'_1, {g_2}^{-1} g''_2 \right)
\delta_{G/H} \left({g'_1}^{-1} L_1, {g''_2}^{-1} L_2 \right) \nonumber \\
&\quad \times 
\Delta \left(g''_1 {g'_2}^{-1} \right) \delta \left({g_1}^{-1} g''_1, {g''_2}^{-1} g'_2 \right)
\delta_{G/H} \left({g''_1}^{-1} L_1, {g'_2}^{-1} L_2 \right)  \ . 
\end{align}
A change of variables
$g''_1 = g'_1 {g''_2}^{-1} g_2$ and $g'_2 = g_2 {g_1}^{-1} g'_1$ 
gives rise to
\begin{align}
& \frac{\kappa^2 v'}{6v} \int dL_1 dL_2 dg_1 dg'_1 dg_2 dg''_2 \; 
\Delta \left(g_1 {g_2}^{-1} \right) \Delta \left(g'_1 {g''_2}^{-1} \right) 
\Delta \left(g'_1 {g''_2}^{-1} g_2 {g'_1}^{-1} g_1 {g_2}^{-1} \right) 
\delta \left(g_1^{-1}g'_1 {g''_2}^{-1} g_2, {g''_2}^{-1} g_2 g_1^{-1}g'_1 \right)
\nonumber \\
&\quad \times
\delta_{G/H} \left(g_1^{-1} L_1, {g_2}^{-1} L_2 \right)
\delta_{G/H}  \left({g'_1}^{-1} L_1, {g''_2}^{-1} L_2 \right)
\delta_{G/H}\left(g_2^{-1} g''_2 {g'_1}^{-1} L_1, {g'_1}^{-1} g_1 g_2^{-1} L_2 \right) \ . 
\end{align}
A further change of variables 
$g_1 g_2^{-1} = \tilde{g}_2^{-1}$ and $g'_1 {g''_2}^{-1} = \tilde{g}'_1$ 
leads to
\begin{align}
&\quad \frac{\kappa^2 v'}{6v} \int dL_1 dL_2 dg_1 d\tilde{g}'_1 d\tilde{g}_2 dg''_2 
\Delta \left(\tilde{g}_2^{-1} \right) \Delta \left(\tilde{g}'_1 \right) 
\Delta \left(\tilde{g}'_1 \tilde{g}_2 g_1 {g''_2}^{-1} \tilde{g'}_1^{-1} \tilde{g}_2^{-1} \right) 
\delta \left(g_1^{-1} \tilde{g}'_1 \tilde{g}_2 g_1, {g''_2}^{-1} \tilde{g}_2 \tilde{g}'_1 g''_2 \right)
\nonumber \\
&\qquad \times
\delta_{G/H}\left(L_1, \tilde{g}_2^{-1} L_2 \right)
\delta_{G/H}\left(L_1, \tilde{g}'_1 L_2 \right)
\delta_{G/H}\left(L_1, \tilde{g}'_1 \tilde{g}_2 g_1 {g''_2}^{-1} \tilde{g'}_1^{-1} \tilde{g}_2^{-1} L_2 \right) \nonumber \\
&= \frac{\kappa^2 v'}{6v} \int dL_1 dL_2 dg_1 d\tilde{g}'_1 d\tilde{g}_2 dg''_2 
\Delta \left(\tilde{g}_2^{-1} \right) \Delta \left(\tilde{g}'_1 \right) 
\Delta \left(g_1 {g''_2}^{-1} \right) 
\delta \left(\tilde{g}'_1 \tilde{g}_2 g_1, g_1 {g''_2}^{-1} \tilde{g}_2 \tilde{g}'_1 g''_2 \right)
\nonumber \\
&\qquad \times
\delta_{G/H}\left(L_1, \tilde{g}_2^{-1} L_2 \right)
\delta_{G/H}\left(L_1, \tilde{g}'_1 L_2 \right)
\delta_{G/H}\left(L_1, g_1 {g''_2}^{-1} L_2 \right) \ .
\end{align}
By making a change of variables $g_1 {g''_2}^{-1} = \tilde{g}$, we obtain
\begin{align}
\label{non-planar_bilocal}
&\quad \frac{\kappa^2 v'}{6v} V \int dL_1 dL_2 d\tilde{g}'_1 d\tilde{g}_2 d\tilde{g} 
\Delta \left(\tilde{g}_2^{-1} \right) \Delta \left(\tilde{g}'_1 \right) 
\Delta \left(\tilde{g} \right)
\nonumber \\
&\qquad \times 
\delta \left(\tilde{g}'_1 \tilde{g}_2 \tilde{g}, \tilde{g} \tilde{g}_2 \tilde{g}'_1 \right) 
\delta_{G/H}\left(L_1, \tilde{g}_2^{-1} L_2 \right)
\delta_{G/H}\left(L_1, \tilde{g}'_1 L_2 \right) 
\delta_{G/H}\left(L_1, \tilde{g} L_2 \right) \ .
\end{align}
Since $\delta \left(\tilde{g}'_1 \tilde{g}_2 \tilde{g}, \tilde{g} \tilde{g}_2 \tilde{g}'_1 \right)  \leq \delta(0) = 1/v$,
\begin{align}
& \mbox{Absolute value of \eqref{non-planar_bilocal}} \nonumber \\
&\leq 
\frac{\kappa^2 v'}{6v} \delta(0) V \int dL_1 dL_2 dh dh' dh''
\left| 
\Delta \left(L_2^{-1} L_1 h \right) \Delta \left(L_2^{-1} L_1 h' \right) 
\Delta \left(L_2^{-1} L_1 h'' \right) 
\right| \nonumber \\
&= \left \{ 
\frac{v V'}{v'} \frac{\lambda N^2}{6V} \frac{V'}{V} \int dL_1 dL_2 dh dh' dh''
\left|
\Delta \left(L_2^{-1} L_1 h \right) \Delta \left(L_2^{-1} L_1 h' \right) 
\Delta \left(L_2^{-1} L_1 h'' \right)
\right| 
\right \} \times \left( \frac{v'}{V'} \right)^2 
\ .
\end{align}
We see that the above quantity is analogous to 
the one (\ref{non-planar diagram in bilocal reduced model})
and suppressed by $(v'/V')^2$ in the $v' \rightarrow 0$ limit compared to
(\ref{planar on G/H}). Thus, the non-planar diagram is suppressed compared to the planar diagram
in both the original and reduced model. 
By comparing (\ref{planar on G}) and (\ref{planar on G/H}), we again obtain
the relation (\ref{relation for free energy}). As in the case of torus, one can show that
(\ref{relation for free energy}) holds to all orders in perturbative expansion.

Defining $\hat{\phi}(g)$ by 
$
\hat{\phi}(g)=e^{iL_A\theta_A}\phi(L)e^{-iL_A\theta_A}
$, 
where $g= e^{i\theta_A t_A}$, we see that
the relation (\ref{relation for correlation functions}) also holds in this case.
Thus, we find that the large-$N$ equivalence in dimensional reduction holds on group manifolds.

Finally, we consider $U(N)$ Yang-Mills theory on $G$:
\begin{equation}
S=\frac{1}{4\kappa^2}\int d^Dx \sqrt{G} G^{AC}G^{BD}\mbox{Tr}(F_{AB}F_{CD}) \ ,
\label{YM on G}
\end{equation}
where $F_{AB}=\partial_A A_B -\partial _B A_A + i[A_A,A_B]$.
We expand the gauge field $A_A$ in terms of the right invariant 1-form as
$
A_M = E_M^A X_A
$.
By using (\ref{Maurer-Cartan equation}), we rewrite (\ref{YM on G}) as
\begin{equation}
S =-\frac{1}{4\kappa^2}
\int d^Dx\sqrt{G} \; \mbox{Tr} \left({\cal L}_AX_B-{\cal L}_BX_A-if_{ABC}X_C+[X_A,X_B] \right)^2 \ .
\label{YM on G 2}
\end{equation}
By imposing 
$
{\cal R}_a X_A = 0  
$ 
on (\ref{YM on G}), we obtain Yang-Mills theory on $G/H$.
The reduced model on $G/H$ is given by
\begin{equation}
S_r= -\frac{v}{v'}\frac{1}{4\kappa^2} \int d^{D-d}\sigma \sqrt{g}\; 
\mbox{Tr} \left({\cal L}_A \tilde{X}_B-{\cal L}_B\tilde{X}_A-i f_{ABC}X_C
+ \left[\tilde{X}_A,\tilde{X}_B \right] \right)^2 \ ,
\label{reduced model of YM on G/H}
\end{equation}
where $\tilde{X}_A(\sigma)=L_A +X_A(\sigma)$.
If $G$ is simple, the gauge theory is massive due to the $f_{ABC} X_C$ term so that
the there is no moduli for the background $\tilde{X}_A=L_A$.
Thus, since the background $\tilde{X}_A=L_A$ is stable, 
the large-$N$ equivalence in dimensional reduction on group manifolds holds. 
Namely, the reduced model is equivalent 
to the original model (\ref{YM on G 2}) in the $N\rightarrow\infty$ 
with $\kappa^2N$ fixed
in the sense that 
(\ref{relation for free energy}) holds and a following relation for Wilson loops \cite{Ishii:2007sy} holds:
\begin{equation}
\left\langle \frac{1}{N}P \exp \left(i\int_0^1d\zeta 
\frac{dx^{M}(\zeta)}{d\zeta} E_{M}^A(x(\zeta))X_A(x(\zeta)) \right) \right\rangle
=\left\langle \frac{1}{N}P \exp \left(i\int_0^1d\zeta 
\frac{dx^{M}(\zeta)}{d\zeta}  E_{M}^A(x(\zeta))\tilde{X}_A(\sigma(\zeta)) 
\right) \right\rangle_r  \ ,
\end{equation}
where $x^{M}(\zeta)$ and $\sigma^{\mu}(\zeta)$ are related through
$g(x) = L(\sigma)h(y)$.

\section{Conclusion and Discussion}
\setcounter{equation}{0}
In this paper, we showed that a theory on a group manifold $G$
is equivalent to the corresponding theory on $G/H$ with $H$ a subgroup 
of $G$ in the large-$N$ limit. 
The degrees of freedom on $G$ are retrieved by the degrees of freedom of matrices in a consistent way with the dimensional reduction to $G/H$.
An advantage of reduction to $G/H$ with
a finite volume compared to reduction to a matrix model is 
that one does not need
to introduce $k$ multiplicity and take the $k\rightarrow\infty$ limit
to extract only planar contribution as in the latter case \cite{Kawai:2009vb,Kawai:2010sf},
since the UV cutoff $V'/v'$ plays the role of extracting planar contribution.
While we showed the equivalence perturbatively, we can show it non-perturbatively based on the continuum Schwinger-Dyson equations as in \cite{Gross:1982at}, 
by assuming the stability of the background, which is a counterpart of the center symmetry.

An interesting application of the large-$N$ equivalence in dimensional reduction on group manifolds
is that the $SU(2|4)$ symmetric
gauge theory on $R\times S^2$ is equivalent to ${\cal N}=4$ super Yang-Mills 
theory on $R\times S^3$ in the large-$N$ limit. (For another
large-$N$ equivalence
between these two theories, see \cite{Ishii:2008ib,Ishiki:2006yr}.)
Both of the theories have gravity
duals, so that the above equivalence would be seen on the gravity side.
It is interesting to search for gravity duals of other large-$N$ equivalences in dimensional reduction \cite{Furuuchi:2005qm,Furuuchi:2005eu,Poppitz:2010bt,Young:2014jma,Shaghoulian:2016xbx}.

\section*{Acknowledgements}

We would like to thank E. Shaghoulian for communication triggering this research
and H. Kawai for comments.
The work of A.T. is supported in part by Grant-in-Aid
for Scientific Research
(No. 15K05046)
from JSPS.



\end{document}